\documentclass[a4paper,11pt]{article}
\pdfoutput=1 

\usepackage{jcappub} 

\usepackage[T1]{fontenc} 
\usepackage{multirow}

\title{\boldmath Revisiting constraints on the photon rest mass with cosmological fast radio bursts}

\author[a,b]{Bao Wang,}
\author[a,b,1]{Jun-Jie Wei,\note{Corresponding author.}}
\author[a,b,1]{Xue-Feng Wu}
\author[c, d, e]{and Mart\'{i}n L\'{o}pez-Corredoira}

\affiliation[a]{Purple Mountain Observatory, Chinese Academy of Sciences,\\Nanjing 210023, China}
\affiliation[b]{School of Astronomy and Space Sciences, University of Science and Technology of China,\\Hefei 230026, China}
\affiliation[c]{Instituto de Astrof\'{i}sica de Canarias, \\E-38205 La Laguna, Tenerife, Spain}
\affiliation[d]{PIFI-Visiting Scientist 2023 of China Academy of Sciences at Purple Mountain Observatory, Nanjing 210023 and National Astronomical Observatories, Beijing 100012, China}
\affiliation[e]{Departamento de Astrof\'{i}sica, Universidad de La Laguna, \\E-38206 La Laguna, Tenerife, Spain}

\emailAdd{jjwei@pmo.ac.cn}
\emailAdd{xfwu@pmo.ac.cn}

\abstract{
Fast radio bursts (FRBs) have been suggested as an excellent celestial laboratory for testing the 
zero-mass hypothesis of the photon. In this work, we use the dispersion measure (DM)--redshift 
measurements of 23 localized FRBs to revisit the photon rest mass $m_{\gamma}$. As an improvement
over previous studies, here we take into account the more realistic probability distributions of DMs
contributed by the FRB host galaxy and intergalactic medium (IGM) from the IllustrisTNG simulation.
To better account for the systematic uncertainty induced by the choices of priors of cosmological 
parameters, we also combine the FRB data with the cosmic microwave background data, the baryon
acoustic oscillation data, and type Ia supernova data to constrain the cosmological parameters 
and $m_{\gamma}$ simultaneously. We derive a new upper limit of $m_{\gamma}\le3.8\times 10^{-51}\;\rm{kg}$, 
or equivalently $m_{\gamma}\le2.1 \times 10^{-15} \, \rm{eV/c^2}$ ($m_{\gamma} \le  7.2 \times 10^{-51} \, \rm{kg}$,
or equivalently $m_{\gamma}\le4.0 \times 10^{-15} \, \rm{eV/c^2}$) at $1\sigma$ ($2\sigma$) confidence level.
Meanwhile, our analysis can also lead to a reasonable estimation for the IGM baryon
fraction $f_{\rm IGM}=0.873^{+0.061}_{-0.050}$. With the number increment of localized FRBs, 
the constraints on both $m_{\gamma}$ and $f_{\rm IGM}$ will be further improved. 
A caveat of constraining $m_{\gamma}$ within the context of the standard $\Lambda$CDM 
cosmological model is also discussed.
}


\arxivnumber{}

\begin{document}
\maketitle
\flushbottom


\section{Introduction}
\label{intro}
It is a widely accepted consensus in modern physics that the rest mass of the photon should be exactly zero.
However, several theories suggest otherwise, such as 
the de Broglie-Proca theory \citep{De1922, Proca1936}, the model of massive photons as dark energy \citep{Kouwn2016}, 
massive photons within an alternative gravity scenario \citep{Bartlett2011}, and other new ideas 
in classical electrodynamics with massive photons \citep{Roscoe2006, Cumalat2011, Spallicci2021}.
Testing the correctness of the photon zero-mass hypothesis has emerged as an attractive task 
in modern physics, as it could potentially shed light on these alternative theories.

Some measurable effects have been predicted when photons have nonzero mass.
The photon rest mass can then be effectively constrained by seeking such effects 
\citep{Goldhaber1971, Lowenthal1973, 1997ASTPS...4.....Z,Tu2005, Okun2006, Goldhaber2010, Spavieri2011}.
Over the last few decades, a number of methods have been proposed to constrain the photon mass, 
including tests of Coulomb's inverse square law \citep{Williams1971}, tests of Amp\`{e}re's law 
\citep{Chernikov1992}, experiments of Cavendish torsion balance \citep{Luo2003}, exploration on 
the gravitational deflection of electromagnetic waves \citep{Lowenthal1973, Accioly2004}, measurement of 
Jupiter's magnetic field \citep{Davis1975}, 
analysis of cosmic magnetic field vector potential \cite{Chibisov1976,Lakes1998}, 
measurement of redshift/blueshift associated with alternative 
fits of the Hubble diagram of type Ia supernovae (SNe Ia; \cite{Spallicci2021, Lopez-Corredoira2022}), 
and so on.

The most direct way to determine the photon mass or, more precisely, place an upper bound on 
the photon mass is to measure the frequency dependence in the velocity of light.
When laboratory experiments suffer from limitations, astrophysical observations afford excellent conditions for higher precision measurements on the relative velocities of electromagnetic waves with different frequencies \citep{Wu2016,Bonetti2016,Bonetti2017,Shao2017,Xing2019,Wei2020,Wang2021,Chang2023,Lin2023,Lovell1964,Warner1969,
Schaefer1999,2016JHEAp..11...20Z,2017RAA....17...13W,2018JCAP...07..045W,2021PhRvD.104j3516B}.
If the photon rest mass is nonzero, then the speed of light in vacuum is no longer a constant but is 
a function of frequency. Two photons with different frequencies, if emitted simultaneously from the same 
astrophysical source and traveling through the same distance, would thus be observed at different times
 (see Ref. \cite{2021FrPhy..1644300W} for a recent review).
As a new kind of millisecond radio transient occurring at cosmological distances, fast radio bursts (FRBs) 
offer the current best astrophysical laboratory for testing the photon mass through the dispersion method 
\citep{Wu2016,Bonetti2016,Bonetti2017,Shao2017,Xing2019,Wei2020,Wang2021,Chang2023,Lin2023}.

While stringent photon mass limits from FRBs have been obtained, some obstacles to 
the pace of advance remain.
The dispersion measure (DM) contribution from the FRB host galaxy ($\mathrm{DM_{host}}$) is not well known, 
and it also degenerates with the DM contributed by the intergalactic medium (IGM; $\rm{DM_{IGM}}$).
Due to the large IGM fluctuation, $\rm{DM_{IGM}}$ at a given redshift can vary significantly along 
different lines of sight \citep{McQuinn2014}. In most previous works, $\mathrm{DM_{host}}$ for all FRB 
host galaxies was treated as an unknown constant, and a global systematic uncertainty was introduced to 
account for the diversity of host galaxy contribution and the large IGM fluctuation. This treatment
amounts to assuming that the probability distributions of $\rm{DM_{IGM}}$ and $\mathrm{DM_{host}}$ are symmetric.
However, numerical simulations show that the distributions of $\rm{DM_{IGM}}$ and $\mathrm{DM_{host}}$ are asymmetric
and have a significant positive skew \citep{Macquart2020,Zhang2020,Zhang2021}. A more appropriate way to model
$\rm{DM_{IGM}}$ and $\mathrm{DM_{host}}$ is thus required in FRB photon mass limits.
Recently, Lin et al. \cite{Lin2023} constrained the photon mass by considering the more realistic 
non-Gaussian distributions for $\rm{DM_{IGM}}$ and $\mathrm{DM_{host}}$, but ignoring the degeneracy 
between the photon mass and cosmological parameters. It should be emphasized that
just relying on the FRB data is difficult to constrain 
the photon mass and cosmological parameters simultaneously, even with only cosmological parameters \cite{Walters2018}.
All previous works neglected the parameter degeneracy and set
the cosmological parameters as fixed values. In such cases, the uncertainties of cosmological parameters might affect 
the constraints on the photon mass.

In this paper, we use the most up-to-date sample of localized FRBs to revisit the photon mass, by properly taking
into account the accurate probability distributions of $\rm{DM_{IGM}}$ and $\mathrm{DM_{host}}$ derived from 
the IllustrisTNG simulation \citep{Zhang2020,Zhang2021}. Compared with previous works, in addition to the photon mass,
the cosmological parameters as well as the baryon fraction in the IGM ($f_{\rm IGM}$) are also treated as free parameters.
In order to break the parameter degeneracy, we combine the DM--$z$ measurements 
from FRBs with the cosmic microwave background (CMB) data from the latest Planck release
\citep{Chen2019,2020A&A...641A...6P}, the baryon acoustic oscillation (BAO) data \citep{Ryan2019}, and 1048 
Pantheon SN Ia data \citep{Scolnic2018}. With the observational data, we constrain the photon mass and model parameters simultaneously. 

This paper is organized as follows. In Section~\ref{sec:method}, we introduce a theoretical framework for 
estimating the probability distributions of $\mathrm{DM_{IGM}}$ and $\mathrm{DM_{host}}$, and a method to 
constrain the photon mass. The observational data and the resulting constraints on the photon mass and 
other model parameters are presented in Section~\ref{sec:Results}. Finally, discussion and conclusions 
are drawn in Section~\ref{sec:Conclusions}.

\section{Theoretical Framework}
\label{sec:method}
\subsection{Dispersion from the Plasma}
\label{sec:Plasma}
FRB signals at low frequencies would travel through the plasma slower than those at high frequencies.
This phenomenon is known as the dispersion effect. The arrival time difference between low- and high-frequency 
photons propagating from a distant source to the observer is described as \citep{2017AdSpR..59..736B,2018JCAP...07..045W,Wei2020}
\begin{equation}\label{tDM}
	\Delta t_{\rm{DM}}
	=\frac{e^{2}}{8 \pi^{2} m_{e} \epsilon_{0} c}\left(\nu_{l}^{-2}-\nu_{h}^{-2}\right)\mathrm{D M_{astro}}\;,
\end{equation}
where $e$ and $m_e$ are the charge and mass of an electron, respectively,
$\epsilon_0$ is the permittivity of vacuum, $c$ is the speed of light in vacuum, and $\nu_{l}$ and $\nu_{h}$ 
are the low and high frequencies of photons, respectively. Here $\mathrm{DM_{astro}}$ 
is the DM contribution from the plasma, which is defined as the integral of electron number density $n_e$ 
along the line of sight.

For an extragalactic FRB, the $\mathrm{DM_{astro}}$ can be divided into four main components:
\begin{equation}\label{DMplasma}
	\mathrm{DM_{astro}}=\mathrm{DM_{ISM}^{MW}}+\mathrm{DM_{halo}^{MW}}+\mathrm{DM_{IGM}}(z)
	+\frac{\mathrm{DM_{host,0}}}{1+z}\;,
\end{equation}	
where $\mathrm{DM_{ISM}^{MW}}$ and $\mathrm{DM_{halo}^{MW}}$ are contributed by the Milky Way's interstellar medium (ISM) and halo, respectively.
$\mathrm{DM_{IGM}}$ represents the IGM portion 
of DM, and $\mathrm{DM_{host,0}}$ is the DM contribution from the host galaxy. The $(1+z)$ factor 
corrects the local $\mathrm{DM_{host,0}}$ to the measured value $\mathrm{DM_{host}}$ \citep{Deng2014}.

The average $\mathrm{D M_{IGM}}$ is related to the cosmological distance scale and the ionization 
fractions of intergalactic hydrogen and helium. If both hydrogen and helium are completely ionized 
(valid below $z\sim3$), then it can be calculated as \citep{2003ApJ...598L..79I,2004MNRAS.348..999I}
\begin{equation}\label{DMIGM}
	\left\langle\mathrm{DM}_{\mathrm{IGM}}(z)\right\rangle  =\frac{21 c \Omega_{b0} H_{0} f_{\mathrm{IGM}}}{64 \pi G m_{p}}H_e(z)\;,
\end{equation}
where $H_0$ is the Hubble constant, $m_p$ is the proton mass, $G$ is the gravitational constant, $\Omega_{b0}$ 
is the present-day baryon density parameter, and $f_{\mathrm{IGM}}$ is the baryon fraction in the IGM.
Although some numerical simulations and observations suggested that $f_{\mathrm{IGM}}$ is slowly 
evolving with redshift \cite{Meiksin2009,Shull2012,Walters2019}, a recent study indicated that such evidence 
is absent in the relatively small amount of FRB data \cite{Wang2022}. Here we thus consider $f_{\mathrm{IGM}}$ as a constant.
Adopting the flat $\Lambda$CDM cosmological model, the dimensionless redshift function
$H_e(z)$ is given by
\begin{equation}\label{He}
	 H_e(z)=\int_{0}^{z} \frac{1+z^{\prime} }{\sqrt{\Omega_{m0}(1+z^{\prime})^3 + 1-\Omega_{m0}} } d z^{\prime}\;,
\end{equation}
where $\Omega_{m0}$ is the matter density parameter.

\subsection{Dispersion from a Nonzero Photon Mass}
\label{sec: Nonzero}
The relative time delay between low- and high-frequency photons caused by the nonzero photon mass 
can be approximatively expressed as \citep{Wu2016,Wei2020}
\begin{equation}\label{tm}
	 \Delta t_{m_{\gamma}} =\frac{1}{2 H_{0}}\left(\frac{m_{\gamma} c^{2}}{h}\right)^{2}\left(\nu_{l}^{-2}-\nu_{h}^{-2}\right) H_{\gamma}(z)\;,
\end{equation}
where $h$ is the Planck constant, and $H_{\gamma}(z)$ is the other dimensionless function,
\begin{equation}\label{Hgamma}
	 H_{\gamma}(z)=\int_{0}^{z} \frac{(1+z^{\prime})^{-2} }{\sqrt{\Omega_{m0}(1+z^{\prime})^3 + 1-\Omega_{m0}}} d z^{\prime}\;.
\end{equation}

It is evident that $\Delta t_{\rm{DM}}$ and $\Delta t_{m_{\gamma}}$ have a similar frequency dependency 
($\Delta t \propto \nu^{-2}$). Thus, we can define an ``effective DM'' arises from a nonzero photon mass
\citep{Shao2017},
\begin{equation}\label{DMgamma}
	 \mathrm{DM}_{\gamma}(z) = \frac{4 \pi^{2} m_{e} \epsilon_{0} c^{5}}{h^{2} e^{2}} \frac{H_{\gamma}(z)}{H_{0}} m_{\gamma}^{2}\;.
\end{equation}
Comparing the two redshift functions $H_e(z)$ and $H_{\gamma}(z)$, one can see that 
the DM contributions from the IGM and a nonzero photon mass ($\mathrm{DM_{IGM}}$ and $\mathrm{DM}_{\gamma}$) 
have different redshift dependencies. Therefore, the dispersion degeneracy between $\mathrm{DM_{IGM}}$ and 
$\mathrm{DM}_{\gamma}$ could in principle be broken when a large redshift measurements of FRBs are available
\citep{Bonetti2016,Bonetti2017,2017AdSpR..59..736B,Shao2017,Wei2020}. 

\subsection{Methodology}
Suppose that the nonzero photon mass effect exists, the observed DM extracted from the dynamical spectrum of
an FRB should include both $\mathrm{DM_{astro}}$ and $\mathrm{DM}_{\gamma}$, i.e.,
\begin{equation}\label{DMobs}
    \mathrm{DM_{obs}}=\mathrm{DM_{astro}}+\mathrm{DM}_{\gamma} \,.
\end{equation}
If the different DM contributions in Equation~(\ref{DMplasma}) can be estimated properly, then we can effectively
identify $\mathrm{DM}_{\gamma}$ from $\mathrm{DM_{obs}}$, thereby providing a constraint on the photon mass $m_{\gamma}$.

The DM due to the Milky Way's ISM, $\mathrm{DM_{ISM}^{MW}}$, can be well estimated based on the Galactic 
electron density models. Here we adopt the NE2001 model to estimate $\mathrm{DM_{ISM}^{MW}}$ \citep{Cordes2002}.
The $\mathrm{DM_{halo}^{MW}}$ term contributed by the Milky Way's halo is not well known but is expected to be
50--80 $\mathrm{pc\;cm^{-3}}$ \citep{Prochaska2019a}. According to this estimation, we consider a Gaussian prior 
on $\mathrm{DM_{halo}^{MW}}=65\pm15$ $\mathrm{pc\;cm^{-3}}$ \citep{Wu2022}.

For a well-localized FRB, the average $\langle\mathrm{DM}_{\mathrm{IGM}}\rangle$ can be calculated using 
Equation~(\ref{DMIGM}). However, due to the density fluctuations of large-scale structure, the real
value of $\mathrm{DM}_{\mathrm{IGM}}$ would vary significantly around the average. Theoretical treatments of 
the IGM and galaxy halos show that the probability distribution for $\mathrm{DM}_{\mathrm{IGM}}$ 
can be well modelled by a quasi-Gaussian function with a long tail \citep{McQuinn2014,Prochaska2019a,Macquart2020},
\begin{equation}\label{PIGM}
	P_{\mathrm{IGM}}\left({\rm DM_{IGM}} \right)=A \Delta^{-\beta} \exp \left[-\frac{\left(\Delta^{-\alpha}-C_{0}\right)^2}{2 \alpha^{2} \sigma_{\mathrm{DM}}^{2}}\right],  \; \Delta>0 \;,
\end{equation}
where $\Delta \equiv \mathrm{DM}_{\mathrm{IGM}}/\langle \mathrm{DM_{IGM}}\rangle$, $A$ is a normalization
coefficient, $\sigma_{\mathrm{DM}}$ is an effective deviation arises from the inhomogeneities of the IGM, 
and $C_0$ is chosen such that the average $\left\langle \Delta \right\rangle=1$. $\alpha$ and $\beta$ are 
two indices related to profile scales, which are set to be $\alpha=\beta=3$ \citep{Macquart2020}.
Using the state-of-the-art IllustrisTNG simulation \citep{Springel2018}, Zhang et al. \cite{Zhang2021} estimated the best-fits of $A$, $\sigma_{\mathrm{DM}}$, and $C_0$ at different redshifts.
Since the $\mathrm{DM}_{\mathrm{IGM}}$ distributions derived from the IllustrisTNG simulation are given in 
discrete redshifts \citep{Zhang2021}, we extrapolate their parameter values to the redshifts of our localized 
FRBs through the cubic spline interpolation method.

The host contribution $\mathrm{DM_{host}}$ is hard to determine, and may change significantly from 
hosts to hosts. Based on the observations of FRB hosts, Zhang et al. \cite{Zhang2020} selected plenty of simulated 
galaxies with similar properties to observed ones from the IllustrisTNG simulation for calculating the $\mathrm{DM_{host}}$ distribution of FRBs. It has been shown 
that the probability distribution of $\mathrm{DM_{host}}$ follows a lognormal function 
\citep{Macquart2020,Zhang2020},
\begin{equation}\label{Phost}
	P_{\mathrm{host}}\left(\mathrm{DM}_{\mathrm {host }} \right)=\frac{1}{\sqrt{2 \pi}\mathrm{DM}_{\mathrm {host }} \sigma_{\mathrm {host }} } \exp \left[-\frac{\left(\ln \mathrm{DM}_{\mathrm {host }}-\mu_{\mathrm {host}} \right)^2}{2 \sigma_{\mathrm {host}}^{2}}\right] \,,
\end{equation}
where $\mu_{\mathrm{host}}$ and $\sigma_{\mathrm {host}}$ denote the mean and standard deviation of $\ln \mathrm{DM}_{\mathrm {host}}$,
respectively. According to the properties of their host galaxies, Zhang et al. \cite{Zhang2020} divided FRBs into three types.
For each type, the best-fitting values of 
$\mu_{\mathrm{host}}$ and $\sigma_{\mathrm {host}}$ at different redshifts are available in their article. 
In our analysis, we use their results to get $\mu_{\mathrm{host}}$ and $\sigma_{\mathrm {host}}$ at any given 
redshifts by cubic spline interpolation. 
It is obvious that different host types correspond to
different $\mathrm{DM_{host}}$ distributions \cite{Zhang2020}. While Lin et al. \cite{Lin2023} used the same $\mathrm{DM_{host}}$ distribution to describe all FRB hosts.

If a sample of localized FRBs is observed, the joint likelihood function is given by \citep{Macquart2020}
\begin{equation}
	\mathcal{L}_{\rm FRB}=\prod_{i=1}^{N_{\mathrm{FRBs}}} P_{i}\left(\mathrm{DM}_{\mathrm{E},i}\right),
\end{equation}
where $N_{\mathrm{FRBs}}$ is the number of available FRBs and $P_{i}\left(\mathrm{DM}_{\mathrm{E}, i}\right)$ 
represents the probability of the observed $\mathrm{DM_{obs}}$ corrected for our Galaxy:
\begin{equation}
\mathrm{DM_{E}}\equiv\mathrm{DM_{obs}}-\mathrm{DM_{ISM}^{MW}}
-\mathrm{DM_{halo}^{MW}}=\mathrm{DM_{IGM}}+\mathrm{DM_{host}}+\mathrm{DM_{\gamma}}\;.
\end{equation}
For an FRB at a given $z_i$, we can calculate $P_{i}(\mathrm{DM}_{\mathrm{E}, i})$ through:
\begin{equation}\label{eq:P_FRB}
	\begin{split}
	P_{i}\left(\mathrm{DM}_{\mathrm{E}, i} \right) &=
	\int_{0}^{\mathrm{DM}_{\mathrm{E}, i}-\mathrm{DM_{\gamma}}}P_{\text {host}}\left(\mathrm{DM}_{\mathrm {host}}\right)  \\
	&\times
	P_{\mathrm {IGM}}\left(\mathrm{DM}_{\mathrm{E}, i}-\mathrm{DM}_{\mathrm {host}}-\mathrm{DM_{\gamma}}\right) \mathrm{dDM}_{\mathrm {host}}\;,
	\end{split}
\end{equation}
where $P_{\mathrm {IGM}}(\mathrm{DM_{IGM}})$ and $P_{\text {host}}(\mathrm{DM_{host}})$ are the probability 
density functions for $\mathrm{DM_{IGM}}$ and $\mathrm{DM_{host}}$, respectively. 

\begin{table}
	\centering
	\caption{Properties of 23 FRBs with redshift measurements.}
	\label{tab:localized FRBs}
	\renewcommand{\arraystretch}{1}
	\begin{tabular}{lccccc} 
		\hline
		Name & Redshift & $\mathrm{DM_{obs}}$ & $\mathrm{D M_{ISM}^{MW}}$  & References \\
		 &  & $(\mathrm{pc \, cm^{-3}})$ & $(\mathrm{pc \, cm^{-3}})$ & \\
		\hline
					FRB 121102 & 0.19273 & 557  & 188.0 & 1 \\
					FRB 180301 & 0.3304 & 536  & 152.0 &  2 \\
					FRB 180916 & 0.0337 & 348.76  & 200.0 & 3\\
					FRB 180924 & 0.3214 & 361.42  & 40.5 & 4\\
					FRB 181112 & 0.4755 & 589.27  & 102.0 & 5\\
					FRB 190102 & 0.291 & 363.6  & 57.3 & 6 \\
					FRB 190523 & 0.66 & 760.8  & 37.0 & 7 \\
					FRB 190608 &  0.1178 & 338.7  & 37.2 & 8 \\
					FRB 190611 & 0.378 & 321.4  & 57.83 & 9 \\
					FRB 190614 & 0.6 & 959.2 & 83.5 & 10 \\
					FRB 190711 & 0.522 & 593.1 & 56.4 & 9 \\
					FRB 190714 & 0.2365 & 504 & 38.0 & 9 \\
					FRB 191001 & 0.234 & 506.92 & 44.7 & 9 \\
					FRB 191228 & 0.2432 & 297.5 & 33.0 & 2 \\
					FRB 200430 & 0.16 & 380.1 & 27.0 & 9 \\
					FRB 200906 & 0.3688 & 577.8 & 36.0 & 2 \\
					FRB 201124 & 0.098 & 413.52 & 123.2 & 11 \\
                    FRB 210117 & 0.2145 & 730 & 34.4 & 12 \\
					FRB 210320 & 0.27970 & 384.8 & 42 & 12 \\
					FRB 210807 & 0.12927 & 251.9 & 121.2 & 12\\
					FRB 211127 & 0.0469 &234.83 & 42.5 & 12 \\
					FRB 211212 & 0.0715 & 206 & 27.1 & 12 \\
                    FRB 220610A & 1.016 & 1457.624 & 31 & 13 \\
		\hline
	\end{tabular}
\begin{description}
\item References: (1) Chatterjee et al. (2017) \cite{Chatterjee2017}; (2) Bhandari et al. (2022) \cite{Bhandari2022};
(3) Marcote et al. (2020) \cite{Marcote2020}; (4) Bannister et al. (2019) \cite{Bannister2019}; (5) Prochaska et al.
(2019) \cite{Prochaska2019b}; (6) Bhandari et al. (2020) \cite{Bhandari2020}; (7) Ravi et al. (2019) \cite{Ravi2019}; 
(8) Chittidi et al. (2021) \cite{Chittidi2021}; (9) Heintz et al. (2020) \cite{Heintz2020}; (10) Law et al. (2020) \cite{Law2020}; (11) Ravi et al. (2022) \cite{Ravi2022}; 
(12) James et al. (2022) \cite{James2022}; (13) Ryder et al. (2022) \cite{Ryder2022}.
\end{description}
\end{table}

\section{Data and Results}
\label{sec:Results}

Currently, 26 extragalactic FRBs have already been localized. FRB 200110E is the nearest one 
($D\sim 3.6$ Mpc), and its redshift is dominated by the peculiar velocity, rather than by the Hubble flow.
It thus has a negative spectroscopic redshift $z=-0.0001$ \citep{Bhardwaj2021b, Kirsten2022}. 
The observed $\mathrm{DM_{obs}}$ value of FRB 181030 is so small ($\sim 103.396$ $\mathrm{pc\;cm^{-3}}$; 
\cite{Bhardwaj2021}) that it would be reduced to a negative value after subtracting $\mathrm{DM_{ISM}^{MW}}$ 
and $\mathrm{DM_{halo}^{MW}}$. That is, the integral upper limit 
($\mathrm{DM_{E}}-\mathrm{DM_{\gamma}}=\mathrm{DM_{obs}}-\mathrm{DM_{ISM}^{MW}}-\mathrm{DM_{halo}^{MW}}$,
where its $\mathrm{DM_{ISM}^{MW}}$ is 40.16 $\mathrm{pc\;cm^{-3}}$ and $\mathrm{DM_{halo}^{MW}}$ is set to be 
$65\pm15\;\mathrm{pc\;cm^{-3}}$) in Equation~(\ref{eq:P_FRB}) would become negative.
Due to an enormously large DM contributed by its host galaxy, FRB 190520B deviates significantly from the expected
$\mathrm{DM_{IGM}}$--$z$ relation \citep{Niu2022}. 
Therefore, we exclude these three FRBs. 
The remaining 23 FRBs are then used for the following analysis. Table~\ref{tab:localized FRBs} lists
the redshifts, $\mathrm{DM_{obs}}$, and $\mathrm{DM_{ISM}^{MW}}$ of these 23 localized FRBs.

\begin{table}
	\centering
	\caption{The $1\sigma$ constraints on all parameters.  
        Both the $1\sigma$ and $2\sigma$ upper limits on the photon mass $m_{\gamma}$ are displayed.}
	\label{tab:result}
	\renewcommand{\arraystretch}{1.3}
	\begin{tabular}{lc}
		\hline
            Parameters & Values \\
		\hline
		\multirow{2}{*}{$m_{\gamma}\;[10^{-51}\,{\rm kg}]$} & $ \le 3.8\; (1\sigma)$ \\
		   & $ \le 7.2\; (2\sigma)$ \\
            \hline
            $f_{\rm IGM}$ & $0.873^{+0.061}_{-0.050}$ \\
            $\mathrm{DM_{halo}^{MW}}\; [\mathrm{pc\;cm^{-3}}]$ & $\le 60.1$ \\
            $\Omega_{m0}$ & $0.309^{+0.005}_{-0.006}$ \\
            $\Omega_{b0}h^2$ & $0.02246\pm 0.00014$ \\
            $H_0\;[{\rm km \, s^{-1} \, Mpc^{-1}}]$ & $67.77\pm 0.43$ \\
            $M_B$ & $-19.415\pm 0.012$ \\
		\hline
	\end{tabular}
\end{table}

In order to break the degeneracies among model parameters, we combine FRB data with up-to-date cosmological 
probe compilations, including CMB, BAO, and SNe Ia. For the CMB data, we use the derived acoustic scale $l_A$, 
shift parameter $R$, and $\Omega_{b0}h^2$ from the latest Planck analysis of the CMB (TT, TE, EE $+$ lowE) 
\citep{Chen2019,2020A&A...641A...6P}. For BAO, we adopt a combination of 11 BAO measurements from 
Ref. \cite{Ryan2019}. For the SN sample, we use the 1048 Pantheon SNe Ia \citep{Scolnic2018}. 
Please see Ref. \cite{Wang2022} for more details on the exact compilations and the likelihood functions of 
these cosmological probes. The final log-likelihood is thus a sum of the separated likelihoods of FRBs, 
CMB, BAO, and SNe Ia:
\begin{equation}
	\ln \mathcal{L}_{\rm tot}=\ln \mathcal{L}_{\rm FRB}+\ln \mathcal{L}_{\rm CMB}+\ln \mathcal{L}_{\rm BAO}+\ln \mathcal{L}_{\rm SN}\;.
\end{equation}

We estimate all model parameters by maximizing $\mathcal{L}_{\rm tot}$ with MCMC analysis as
implemented by the $emcee$ package in Python \citep{Foreman2013}. In our model, the free parameters include
the cosmological parameters ($\Omega_{m0}$, $\Omega_{b0}h^{2}$, and $H_0$), the SN absolute magnitude $M_B$,
the DM contribution from the Milky Way's halo ($\mathrm{DM_{halo}^{MW}}$), the IGM baryon fraction $f_{\rm IGM}$,
and the photon mass $m_{\gamma}$. In our baseline analysis, we set uniform priors on $\Omega_{m0}\in[0,\;1]$,
$\Omega_{b0}h^{2}\in[0.01,\;0.05]$, $H_0\in[50,\;90]$ $\mathrm{km\;s^{-1}\;Mpc^{-1}}$, $M_B\in[-30,\;0]$,
$f_{\rm IGM}\in[0,\;1]$, and $m_{\gamma}\in[0,\;10^{-49}]$ $\mathrm{kg}$. Given the estimation of 
$\mathrm{DM_{halo}^{MW}}\approx50-80$ $\mathrm{pc\;cm^{-3}}$ \citep{Prochaska2019a}, we set a Gaussian prior on
$\mathrm{DM_{halo}^{MW}}=65\pm15$ $\mathrm{pc\;cm^{-3}}$ and marginalize it over the range of $[50,\;80]$ $\mathrm{pc\;cm^{-3}}$.

\begin{figure}
	\centering
	\includegraphics[width=1.0\textwidth]{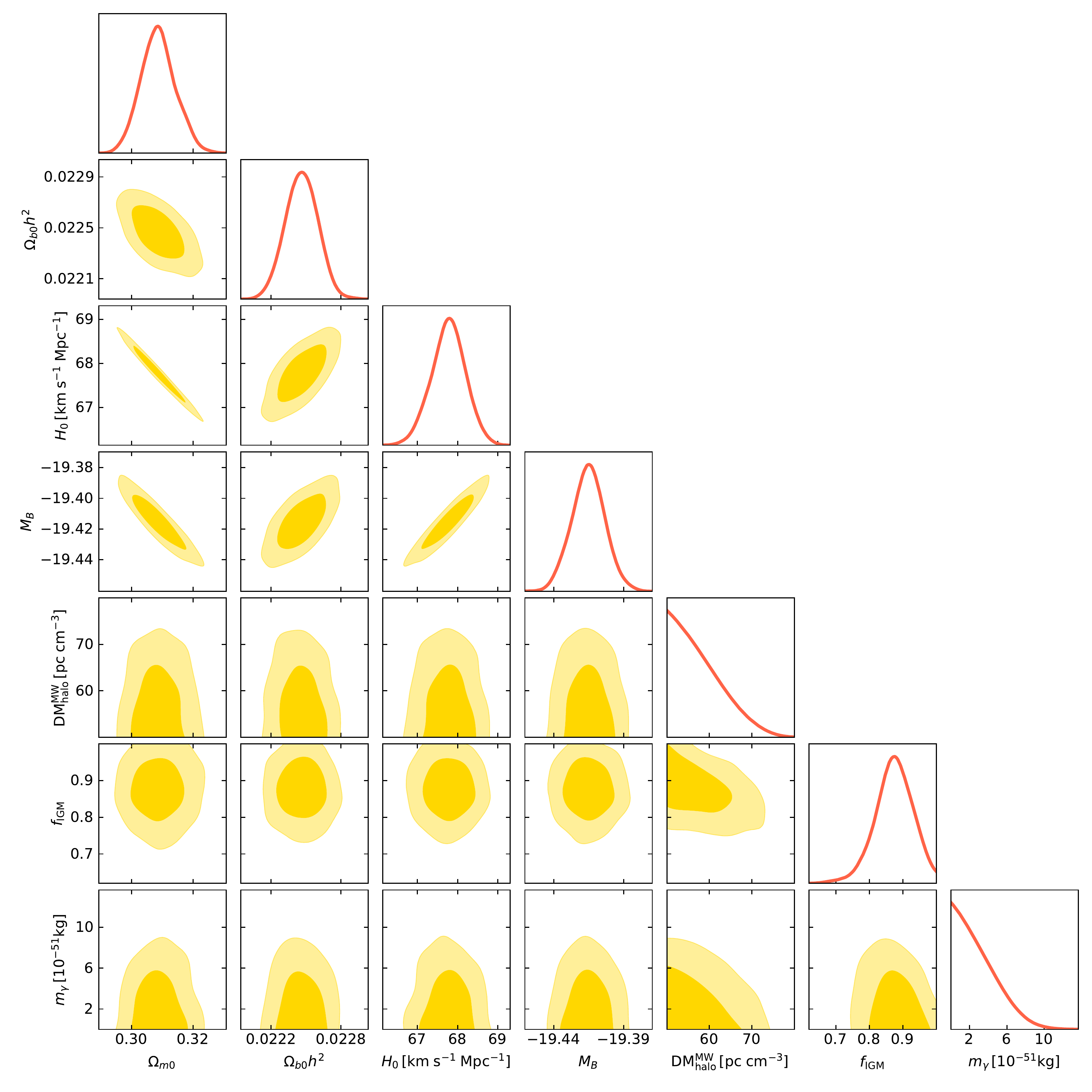}
	\caption{Posterior plots for the parameters $\Omega_{m0}$, $\Omega_{b0}h^{2}$, $H_0$, $M_B$, $\mathrm{DM_{halo}^{MW}}$, $f_{\rm IGM}$, and $m_{\gamma}$, constrained by the FRB, CMB, BAO, and SN Ia data. 
    The contours show the 68\% and 95\% confidence intervals.}
    \label{fig:probability distribution}
\end{figure}

\begin{figure}
	\centering
	\includegraphics[width=0.8\textwidth]{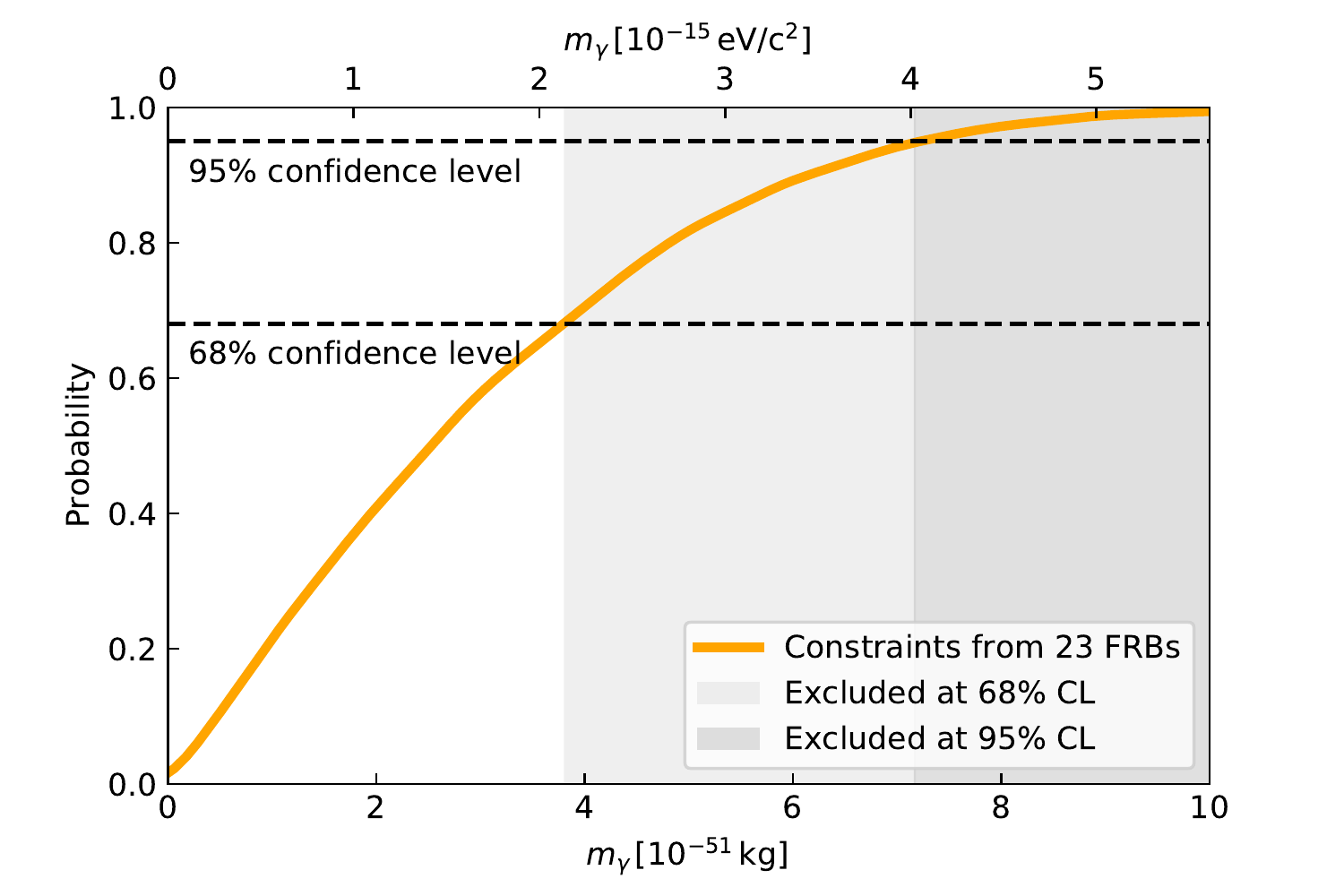}
	\caption{Cumulative posterior distribution for the photon mass $m_{\gamma}$.
	The grey areas mark the excluded values at $1\sigma$ and $2\sigma$ confidence levels.}
    \label{fig:accumulative distribution}
\end{figure}

Using the MCMC method, the marginalized $1\sigma$ constraints on these seven parameters are presented in Table~\ref{tab:result}.
For the photon mass, both the $1\sigma$ and $2\sigma$ upper limits are displayed.
The posterior distributions and the confidence contours for these parameters are plotted in Fig.~\ref{fig:probability distribution}.
We find that the IGM baryon fraction is inferred to be
$f_{\rm IGM}=0.873^{+0.061}_{-0.050}$, which is in good agreement with previous works \citep{Li2020,Wang2022}.
In Fig.~\ref{fig:accumulative distribution}, we show the cumulative posterior distribution of the photon mass $m_{\gamma}$. One can see from this plot that the $1\sigma$ and $2\sigma$ upper limits on $m_{\gamma}$ are
\begin{equation}
	m_{\gamma} \le  3.8 \times 10^{-51} \, \rm{kg} \simeq 2.1 \times 10^{-15} \, \rm{eV/c^2}  
\end{equation}
and 
\begin{equation}
	m_{\gamma} \le  7.2 \times 10^{-51} \, \rm{kg} \simeq 4.0 \times 10^{-15} \, \rm{eV/c^2} \;,
\end{equation}
respectively. Our results do not indicate that photons have definite mass, which means that the massless photon postulate is solid on the basis of the FRB measurements. We note that, compared with other cosmological probes, the FRB data have 
negligible effects on the estimated cosmological parameters. While the constraints on both $m_{\gamma}$ and $f_{\rm IGM}$ 
are drawn primarily from the FRB data.

Table~\ref{tab:overview} presents a summary of upper limits on the photon mass from FRBs.
As an improvement over previous studies, the more realistic probability distributions of $\mathrm{DM_{IGM}}$ 
and $\mathrm{DM_{host}}$ with asymmetric tails to large-DM values have been considered in Wang et al. \cite{Wang2021}
(hereafter W21), Lin et al. \cite{Lin2023} (hereafter L23), and our work. As shown in Table~\ref{tab:overview}, W21 and L23 used a catalog of 129 FRBs
and 17 localized FRBs to constrain the photon mass, yielding $m_{\gamma}\le3.1\times 10^{-51}\;\rm{kg}$ and
$m_{\gamma}\le4.8\times 10^{-51}\;\rm{kg}$, respectively. These constraints are comparable to our result of 
$m_{\gamma}\le3.8\times 10^{-51}\;\rm{kg}$ at $1\sigma$ confidence level, however, it is still useful to compare
differences in the analysis methods. Since most of the FRBs in their catalog have no redshift measurements, 
W21 had to use the observed $\mathrm{DM_{obs}}$ values to estimate the pseudo redshifts. Our photon mass limit 
from 23 localized FRBs is a little looser than that of W21, while avoiding potential bias from the estimation 
of redshift. Furthermore, neither W21 nor L23 accounted for the influence of the uncertainties of cosmological 
parameters on the photon mass limits. Unlike W21 and L23 that fix the cosmological parameters at certain values,
we treat them as free parameters, so the effect caused by the uncertainties of cosmological parameters 
can be well handled in our analysis.

\begin{table}
	\centering
	\small
	\caption{A summary of upper limits on the photon rest mass $m_{\gamma}$ from FRBs.}
	\label{tab:overview}
	\renewcommand{\arraystretch}{1}
	\begin{tabular}{llll}
		 \hline
	Author (Year)	 & Sources & Frequency range & $m_{\gamma}$ ($\mathrm{kg}$) \\
		 \hline
		 Wu et al. (2016) \cite{Wu2016} & FRB 150418 & 1.2--1.5 GHz & $5.2 \times 10^{-50}$ \\
		 Bonetti et al. (2016) \cite{Bonetti2016} & FRB 150418 & 1.2--1.5 GHz & $3.2 \times 10^{-50}$ \\
		 Bonetti et al. (2017) \cite{Bonetti2017} & FRB 121102 & 1.1--1.7 GHz & $3.9 \times 10^{-50}$ \\
   		 Shao \& Zhang (2017) \cite{Shao2017} & 21 FRBs (20 of them without &  $\sim$GHz & $8.7\times 10^{-51}\;(1\sigma)$ \\
                          &  redshift measurement) &              & $1.5\times 10^{-50}\;(2\sigma)$ \\
		 Xing et al. (2019) \cite{Xing2019} & FRB 121102 subpulses & 1.34--1.37 GHz & $5.1 \times 10^{-51}$ \\
		 Wei \& Wu (2020) \cite{Wei2020} & 9 localized FRBs & $\sim$GHz & $7.1 \times 10^{-51}\;(1\sigma)$ \\
                         &                  &           & $1.3 \times 10^{-50}\;(2\sigma)$ \\
          	Wang et al. (2021) \cite{Wang2021} & 129 FRBs (most of them without & $\sim$GHz & $3.1 \times 10^{-51}\;(1\sigma)$ \\
                          &   redshift measurement)       &           & $3.9 \times 10^{-51}\;(2\sigma)$ \\
          	Chang et al. (2023) \cite{Chang2023} & FRB 180916B$^{\rm a}$ & 124.8--185.7 MHz & $6.0 \times 10^{-47}$ \\
          	Lin et al. (2023) \cite{Lin2023} & 17 localized FRBs & $\sim$GHz & $4.8 \times 10^{-51}\;(1\sigma)$ \\
                          &                   &           & $7.1 \times 10^{-51}\;(2\sigma)$ \\
		 This work & 23 localized FRBs & $\sim$GHz & $3.8 \times 10^{-51}\;(1\sigma)$ \\
                   &                   &           & $7.2 \times 10^{-51}\;(2\sigma)$ \\
		\hline
	\end{tabular}
\begin{description}
  \item[\emph{Note.}] {$^{\rm a}$The limit of Ref. \cite{Chang2023} was obtained based on the second-order photon mass effect.}
\end{description}
\end{table}

\section{Conclusions and Discussion}
\label{sec:Conclusions}
Cosmological FRBs have been proposed as an ideal probe for constraining the photon rest mass $m_\gamma$. 
Similar to the dispersion from the plasma, the time delay due to the $m_{\gamma}\neq0$ effect is proportional 
to $\nu^{-2}$ in the first-order approximation. Therefore, a nonzero photon mass can produce an additional 
$\mathrm{DM}_{\gamma}$, which provides a theoretical basis to constrain $m_\gamma$. A key challenge in 
this dispersion method for constraining $m_\gamma$, however, is to distinguish $\mathrm{DM}_{\gamma}$ from 
other DMs contributed by the FRB host galaxy ($\mathrm{DM_{host}}$) and IGM ($\mathrm{DM_{IGM}}$). Moreover,
another problem that hinders such studies is the strong degeneracy between $m_\gamma$ and cosmological parameters.

In this work, we revisited the photon mass from the most up-to-date sample of 23 localized FRBs. Unlike most of the
previous works that assumed the $\mathrm{DM_{host}}$ value as an unknown constant and introduced a systematic 
uncertainty to account for the diversity of host galaxy contribution and the large IGM fluctuation, we considered
the more realistic probability distributions of $\mathrm{DM_{IGM}}$ and $\mathrm{DM_{host}}$ derived from 
the IllustrisTNG simulation. Additionally, all previous studies fixed the cosmological parameters at certain values
and neglected the parameter degeneracy. To account for the systematic uncertainty resulting from the choices of 
priors of cosmological parameters, it would be better to treat the cosmological parameters as free parameters
and infer their values from the observational data. With this aim, we combined the DM--$z$ measurements from
FRBs with the CMB, BAO, and SN Ia data to constrain cosmological parameters and $m_\gamma$ simultaneously.

\begin{figure}
	\centering
	\includegraphics[width=1\textwidth]{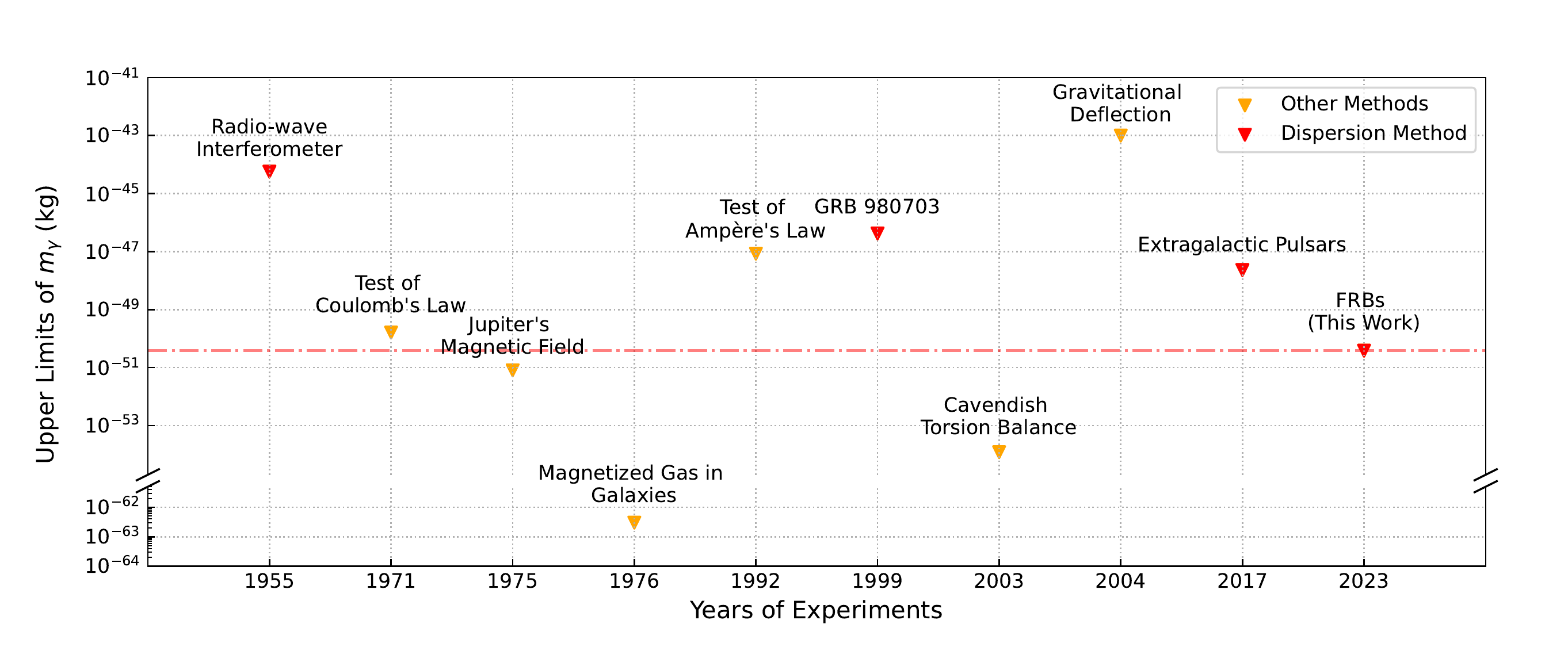}
	\caption{Upper limits on the photon mass from various experimental methods (in temporal order).
    The results obtained through the dispersion method are marked in red, including the limits from the radio-wave interferometer
    ($m_{\gamma}\le5.7\times 10^{-45}\;\rm{kg}$; \cite{Florman1955}), GRB 980703 ($m_{\gamma}\le4.2\times 10^{-47}\;\rm{kg}$;
    \cite{Schaefer1999}) , extragalactic pulsars ($m_{\gamma}\le2.3\times 10^{-48}\;\rm{kg}$; \cite{2017RAA....17...13W}), 
    and FRBs ($m_{\gamma}\le3.8\times 10^{-51}\;\rm{kg}$; this work). The results obtained through other experimental methods
    are marked in orange, including test of Coulomb's law ($m_{\gamma}\le1.6\times 10^{-50}\;\rm{kg}$; \cite{Williams1971}), 
    measurement of Jupiter's magnetic field ($m_{\gamma}\le8\times 10^{-52}\;\rm{kg}$; \cite{Davis1975}), test of 
    Amp\`{e}re's law ($m_{\gamma}\le8.4\times 10^{-48}\;\rm{kg}$; \cite{Chernikov1992}), the experiment of Cavendish torsion 
    balance ($m_{\gamma}\le1.2\times 10^{-54}\;\rm{kg}$; \cite{Luo2003}), analysis of the mechanical stability of the magnetized gas ($m_{\gamma}\le3\times10^{-63}\;\rm{kg}$; \cite{Chibisov1976}), and gravitational deflection of 
    radio waves ($m_{\gamma}\le10^{-43}\;\rm{kg}$; \cite{Accioly2004}).
    The red dot-dashed line corresponds to the $1\sigma$ upper limit of $m_{\gamma}$ from FRBs.}
    \label{fig:lab and astrophysical methods}
\end{figure}

In the light of the above method, we obtained a new upper limit on the photon mass at $1\sigma$ ($2\sigma$) 
confidence level, i.e., $m_{\gamma}\le3.8\times 10^{-51}\;\rm{kg}$, or equivalently 
$m_{\gamma}\le2.1 \times 10^{-15} \, \rm{eV/c^2}$ ($m_{\gamma} \le  7.2 \times 10^{-51} \, \rm{kg}$,
or equivalently $m_{\gamma}\le4.0 \times 10^{-15} \, \rm{eV/c^2}$). Moreover, a reasonable estimation for the IGM baryon
fraction is simultaneously achieved, i.e., $f_{\rm IGM}=0.873^{+0.061}_{-0.050}$, which is well consistent with 
other recent results \citep{Li2020,Wang2022}.
To illustrate the competitiveness of FRB constraints, in Fig.~\ref{fig:lab and astrophysical methods}
we plot the constraints from all various (laboratory and astrophysical) methods. 
As shown in Fig.~\ref{fig:lab and astrophysical methods}, by analyzing the mechanical stability of the magnetized gas 
in galaxies, Chibisov obtained the most stringent limit on the photon mass of $m_{\gamma}\le3\times10^{-60}\,\rm{kg}$ 
\cite{Chibisov1976}. However, Chibisov's approach depends in a critical way on many assumptions, and the reliability of 
this result remains somewhat unclear. One can also see from Fig.~\ref{fig:lab and astrophysical methods} that
the photon mass limit obtained from FRBs is ordinary, but it performs best within the same dispersion method.

In the end, we would like point out a caveat to our method.
Throughout this article, we assume the standard $\Lambda$CDM cosmological model. But massive photons would 
change the cosmological model: for instance, producing non-cosmological redshifts, variations in the propagation 
of light, mass distribution during the history of the Universe, no need of dark energy, etc. In other words, 
our photon mass limits from cosmological FRBs are valid provided that the $\Lambda$CDM model is maintained 
even with a massive photon, which might be not the case if the mass of the photon is significantly different 
from zero.



\acknowledgments
We are grateful to the anonymous referee for his/her helpful comments.
This work is partially supported by the National SKA Program of
China (2022SKA0130100), the National Natural Science Foundation of China (grant No.
12041306), the Key Research Program of Frontier Sciences (grant No. ZDBS-LY-7014)
of Chinese Academy of Sciences, International Partnership Program of Chinese Academy of Sciences
for Grand Challenges (114332KYSB20210018), the CAS Project for Young Scientists in Basic Research
(grant No. YSBR-063), the CAS Organizational Scientific Research Platform for National Major
Scientific and Technological Infrastructure: Cosmic Transients with FAST, the Natural Science
Foundation of Jiangsu Province (grant No. BK20221562), and the Young Elite Scientists
Sponsorship Program of Jiangsu Association for Science and Technology.
M.L.C. is supported by Chinese Academy of Sciences President's International Fellowship Initiative 
(grant No. 2023VMB0001).










\newcommand{\prl}{Phys. Rev. Lett.}
\newcommand{\prd}{Phys. Rev. D}
\newcommand{\apjl}{Astrophys. J. Lett.}
\newcommand{\apj}{Astrophys. J.}
\newcommand{\aj}{Astron. J.}
\newcommand{\mnras}{Mon. Not. Roy. Astron. Soc.}
\newcommand{\aap}{Astron. Astrophys.}
\newcommand{\nat}{Nature}
\newcommand{\jcap}{J. Cosmol. Astropart. Phys.}
\newcommand{\pasp}{Publications of the Astronomical Society of the Pacific}

\bibliography{ref2}{}
\bibliographystyle{JHEP}

\end{document}